\documentclass[12pt]{article} 
\usepackage[usenames, dvipsnames]{color}
\usepackage{graphicx}
\usepackage{fancyhdr}
\usepackage{amsmath}
\begin{document} 

\def\CH{{\cal H}}
\def\caT{{\mathcal T}^{(\al)}}
\def\hk{{\mathcal H}_K}
\def\hak{{\mathcal H}_K^{\al}}
\def\h2ak{{\mathcal H}_K^{(\al)}}
\def\beq{\begin{equation}}
\def\eeq{\end{equation}}
\def\al{\alpha}
\def\be{\beta}
\def\ga{\gamma}
\def\cA{{\mathcal A}}
\def\cB{{\mathcal B}}
\def\cC{{\mathcal C}}
\def\cE{{\mathcal E}}
\def\cF{{\mathcal F}}
\def\cH{{\mathcal H}}
\def\cI{{\mathcal I}} 
\def\cL{{\mathcal L}}
\def\cM{{\mathcal M}}
\def\cM{{\mathcal M}}
\def\cN{{\mathcal N}}
\def\cS{{\mathcal S}}
\def\cT{{\mathcal T}}
\def\cU{{\mathcal U}}
\def\cX{{\mathcal X}}
\def\cY{{\mathcal Y}}

\renewcommand{\marginpar}[2][]{}

\begin{center}
\large{Emanuel Parzen: A Memorial,  and 
a Model With the Two Kernels That He Championed
}
\end{center}
\begin{center}
{\sf {Grace Wahba}\footnote[1]{Research supported 
in part by
NSF Grant DMS-1308847 and a Consortium of NIH Institutes 
under Award Number U54AI117924}\\
Department of Statistics,
Department of Computer Sciences \\
and Department of Biostatistics and Medical Informatics\\
University of Wisconsin, Madison\\
\today 
}
\end{center}
\vspace{0.8in}
\marginpar{absrr}
\begin{abstract}
Manny Parzen passed away in February 2016, and 
this article is written partly as a memorial and  
appreciation. 
Manny made important contributions
to several areas, but the two that influenced me 
most were his  
contributions to kernel density 
estimation and to Reproducing Kernel Hilbert 
Spaces, the two kernels of the title.  
Some fond memories of Manny as a PhD 
advisor begin this memorial, followed by 
a discussion of Manny's influence on 
density estimation and RKHS 
methods. A picture gallery of trips  
comes next, followed by the technical part  
of the article. Here our goal is to show 
how risk models can be built using RKHS penalized
likelihood methods where subjects have personal
(sample) densities
which can be used as {\it attributes}  in such models.

\end{abstract}

\section{Scholar, teacher, friend}
\subsection{Manny as PhD advisor}
\marginpar{scholar}
In 1962 I was a single mom working at a D. C. area think
tank and also working towards a Masters at the University of 
Maryland-College Park when I read {\it Mod Prob}
\cite{parzen:1960} and {\it Stochastic Processes}
\cite{parzen:1962b}, and imagining an impossible dream of 
relocating to the West Coast, going to Stanford and having 
Prof. Emanuel Parzen as a thesis advisor. Well, sometimes
dreams do come true.  I got a new job with a group  
at IBM in D. C. and shortly thereafter, they moved the whole
group to the Bay area. Voil\`{a}, admission to the IBM work 
study program and to Stanford and eventually I became 
Manny's fifth student. Soon we were meeting regularly and 
he would enthusiastically listen to my initial attempts
at research. I have many fond memories of my five years
as a student (1962-66 and postdoc 1967). One of my first
memories is an elegant dinner that Manny and his wife Carol threw
for a group of students - convincing me that 
academic life was something to be desired. Carol was  
always part of things, knowing many of Manny's students
and being enthusiastic about their academic lives. 
Once I got through the first 
year qualifier (at Stanford they had first and second 
year qualifiers) I didn't really worry about making it 
through (sort of).   Manny was always encouraging, and  
he was the most positive, outgoing and 
optimistic person I had ever met. 
Another fond memory is a class Manny taught and held
on nice days on the grass in front of the 
old Sequoia hall, 
later demolished and replaced by
a more modern new Sequoia hall in 1998. 
Manny was one of the major figures in time
series analysis and this class reflected his 
contributions to the field at the time, one  example
being the fundamental paper  \cite{parzen:1961}. 
It was the 
first time I heard him talk about Reproducing Kernel
Hilbert Spaces \cite{aronszajn:1950}, although he had 
published several papers utilizing them around that 
time, for example \cite{parzen:1961b}
\cite{parzen:1963a}.
In any case, 
some of his elegant approaches to RKHS remained dormant
in my brain for a few years (more on that later) and 
I went on to write a dissertation on vector valued
time series \cite{wahba:1968, wahba:1969a}
under his supervision. Manny looked after his students. 
It turned out that E. J. (Ted) Hannan, in Canberra, 
Australia was one of Manny's scientific correspondents and was 
working on something similar to my thesis work. 
Manny sent him  what was to become \cite{wahba:1969a}.
Recalling that in the 60's it could take three weeks
to get a package to Australia, it happened that 
Hannan sent Manny what was to become \cite{hannan:1967}
and the manuscripts crossed in the mail - a bit different
than instant communication around the world today. 
I think Manny had written Hannan about my work
along with sending
the manuscript, and although Hannan's paper ultimately
was published several years before mine, he generously
mentioned my work in his paper. I received the PhD in June 
of 1966 and, if memory serves,  Manny took me 
and my Dad, who had come from New Jersey for
the graduation, to lunch at the faculty club. 
I went on to spend a year
as a postdoc with Manny. During that year he apparently
contacted a number of his friends, 
including George Box at Madison, resulting in 
a large number of invitations to give a lecture, 
and, ultimately
eight job offers. The late 60's were a good time to be 
looking for an academic job, as universities were growing
to accommodate the children of the veterans returning from 
the Second World War. The process was much simpler, too - 
Manny made a bunch of phone calls, I gave some talks, and 
eventually I got a letter with a short paragraph saying something 
like ``We would like to offer you a position as an assistant
professor with the academic year salary of (say) \$10,000.
Please let us know by such-and-such a date whether you 
accept". Today the successful applicant will get a large
packet with enough rules and regulations to keep busy 
reading them for a week. Not to mention
the application process whereby the potential hire is usually
responding to a job posting, enters
a large number of documents into an on line website, 
while the applicant's  references
have to enter detailed information 
into another website. 
In September of 1967 I left sunny California
for the frozen winters of the Midwest, the University of 
Wisconsin-Madison, where there was a powerful numerical analysis
group and a fertile place to nurture the seeds of function
estimation in RKHS.

\subsection{Manny's influence -- density estimation}\marginpar{madison}
  Manny had written a fundamental paper on density estimation
  \cite{parzen:1962d} (more on this in the 
  technical section below)
  and being aware of this work led me to write a bunch of 
  papers on density estimation and spectral density 
  estimation, including 
  \cite{wahba:1975c} 
  \cite{wahba:1977b}  
  \cite{wahba:1980c} 
  \cite{wold:wahba:1975e}. 
In the early seventies there was a lot of 
 discussion about tuning
 nonparametric models of various kinds.   Manny contributed
 to this issue in the context of time series, in 
 his CATS tuning criteria 
 \cite{parzen:1974}, and in his influence on those around 
 him,  for example, the last three papers above.

\subsection{Manny's influence --  RKHS}
   In 1967, when I arrived in Madison, 
   there was a large 
   group of staff and visitors working excitedly in 
   numerical analysis and approximation theory.
   They were  members of  
   the Mathematics Research Center which was located in 
   Stirling Hall,  
   the building  that was later blown up in 
   August of 1970 in 
   protest against the Vietnam war.  I had a part time
   appointment there along with my position in the 
   Statistics Department. Leading researchers
   in approximation theory and numerical analysis
   were there, including I. J. Schoenberg, 
   Carl deBoor, Larry Schumaker, Zuhair Nashed 
   and others. There  
   was much interest in splines, the first of which 
   were invented by Schoenberg in the forties. 
   Tea was served mid morning and mid afternoon accompanied
   by lively discussion. 
   
   In the midst of this creativity, 
   that brain space holding memories of RKHS from 
   Manny's class perked up, and George Kimeldorf, 
   who was a visitor to the MRC at the time, and I 
   together realized that we could derive 
   Schoenberg's polynomial smoothing spline as an 
   optimization problem in an RKHS, and moreover the abstract
   structure for doing that was highly generalizable. 
   We went on to produce three papers together about 
   RKHS \cite{kimeldorf:wahba:1970a} 
   \cite{kimeldorf:wahba:1970b} 
   \cite{kimeldorf:wahba:1971},
   this last paper  giving a closed form expression for the 
   solution to the penalized likelihood optimization 
   problem, where the penalty is a square norm or 
   seminorm in an RKHS - the representer theorem. 
   It was accepted within three
   weeks, something that I never experienced again. 
   RKHS methods seemed to occupy a small niche until
   around 1996 when it became widely known that the 
   Support Vector Machine (SVM), much appreciated by computer
   scientists for its classification prowess,
   could be obtained as the solution to an optimization
   problem in an RKHS. More on this story can be found in 
   \cite{wahba:2013} pp. 486--495. 
   Lin {\it et al}  \cite{lin:wahba:zhang:lee:2002} showed that 
   the SVM was estimating the sign of the log 
   odds ratio, and copious applications of RKHS 
   methods are now part of computer scientists' and 
   statisticians' toolkits. 
\section{Trips} \marginpar{trips1}
In 1984, Hirotugu Akaike threw a fun conference in 
Tokyo focused mostly around a group that was interested
in time series and other common interests of 
Akaike and Manny. There were
exciting sightseeing trips and social events almost 
every evening. I'm sure things are different now, 
but the gracious ladies of the office staff served 
us tea and little cakes on conference days. One evening 
when I surprisingly didn't see anything on the schedule,  
one of Akaike's younger (and presumably single) colleagues
asked me if he could take me out to dinner, and 
I was quite charmed. I didn't stop to wonder where
everyone else was, but some years later I decided
that it must have been planned to allow
the men to visit some place that didn't
expect women e. g. a  geisha teahouse, but I'll never know.  
\newpage
Figure 1 is a picture from the Tokyo trip. 
\begin{figure}[h]\label{pics/tokyo}
\begin{center}
\includegraphics[width=.70\textwidth]{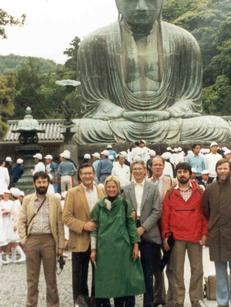}
\caption{Akaike Time Series Conference, Tokyo 1984. l. to r. 
Victor Solo, Manny, me, Wayne Fuller, Bill Cleveland,
Bob Shumway, David Brillinger}
\end{center}
\label{pics/tokyo}
\end{figure}

\newpage
In 1989 there was  a swell sixtieth  \marginpar{fig2} 
birthday party for
Manny, including scientific colleagues, Texas A\&M 
bigwigs and Parzen family. Everyone had a ball, and 
Figure 2 is a scene from the party - there is 
Manny outgoing and smiling as ever. 
\begin{figure}[htb]
\begin{center}
\includegraphics[width=.80\textwidth]{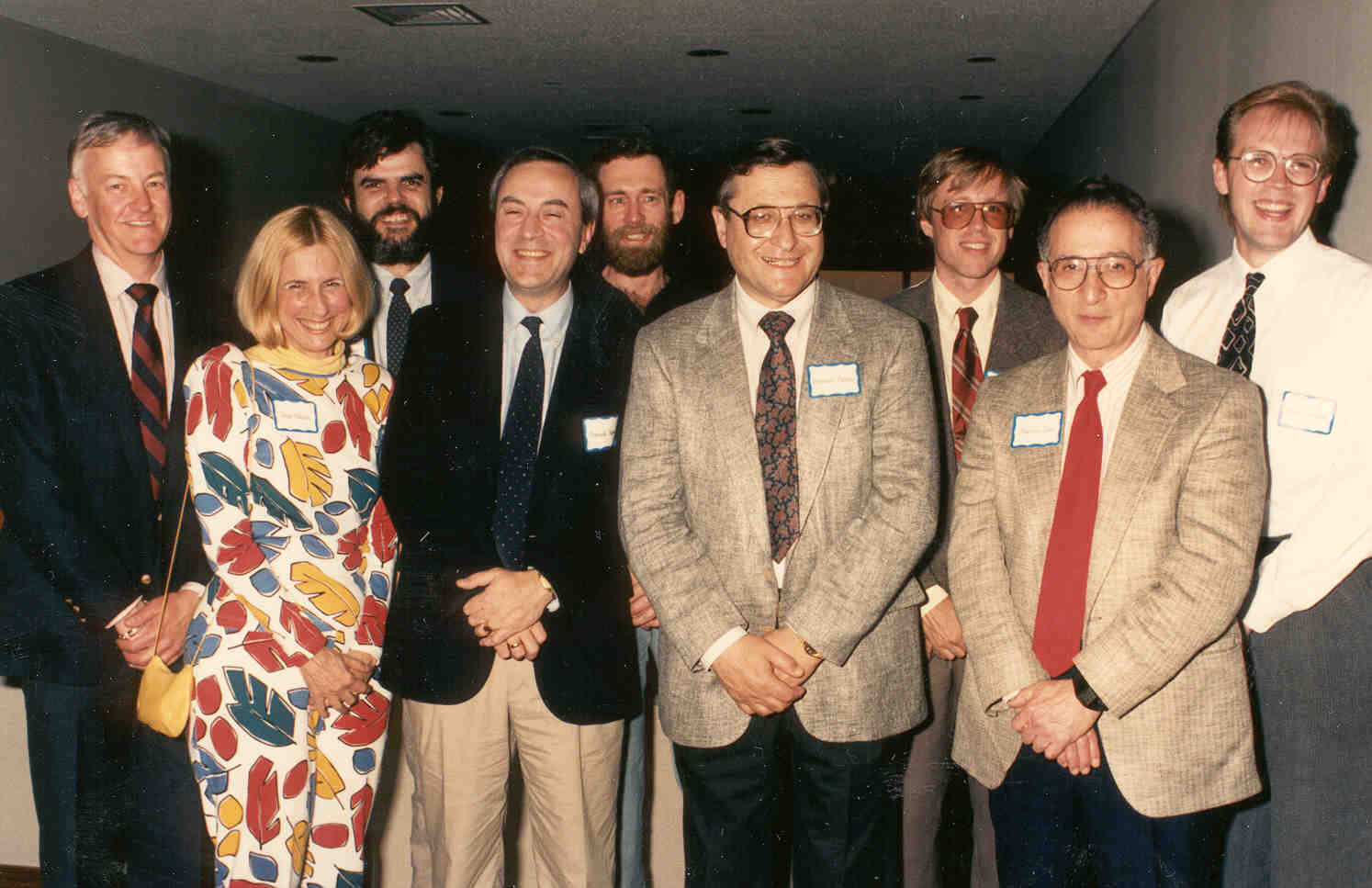}
\caption{Manny's 60th Birthday, 1989, College Station, TX.
l. to r. Don Ylvisaker, me, Joe Newton, 
Marcello Pagano,  
Randy Eubank,
Manny, Will Alexander, Marvin Zelen, Scott Grimshaw
}
\end{center}
\label{pics/60bday}
\end{figure}

\newpage
Marvin Zelen and I attended the  JSM2005 
Gottfried  Noether 
Scholars Award and are congratulating the
winners, Manny and Gerda Claeskins,  in Figure 3. 
\begin{figure}[htb]
\begin{center}
\includegraphics[width=.80\textwidth]{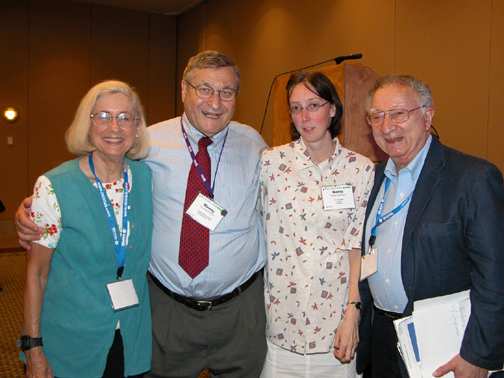}
\caption{At the Gottfried Noether Senior and 
Junior Researchers Awards Ceremony, JSM 2005, 
to Manny Parzen and 
Gerda Claeskens. l. to r. me, Manny, Gerda, 
Marvin Zelen}
\end{center}
\label{pics/claeskens}
\end{figure}

\newpage
Manny was the featured speaker at the 
The Pfizer Colloquium 2006 at UConn, 
with Joe Newton and myself as discussants. 
Joe and I sat for 
a ``Conversation with Many Parzen", 
(see Figure 4)
which was videotaped, and the main thing 
I remember about that was the fact that 
the video was recording off the cuff remarks 
and I was afraid of making a dumb one. 
Manny is smiling as usual but I look a bit tense. 

\begin{figure}[htb]
\begin{center}
\includegraphics[width=.80\textwidth]{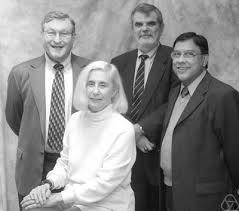}
\caption{Manny, me, Joe Newton, Nitis Mukhopadhyay at 
the Pfizer Colloquium 2006 in Manny's honor at UConn.
}
\end{center}
\label{pics/newton}
\end{figure}

\newpage
\section{Manny, a man of many interests}     \marginpar{sec2}
  Manny had a major role in a number of fundamental
  areas in the development of the Statistical 
  Canon. Aside from density estimation and 
  RKHS, these
  include time series modeling, spectral
  density estimation, and in later years, 
  quantile estimation. However, in this chapter
  we will limit ourselves to Parzen window 
  density estimation and RKHS,  two of Manny's
  areas I have worked in. Interestingly  
  Manny's work is fundamental to 
  the two different kinds of kernels that have 
  played  important  roles in the development 
  of modern statistical methodology.
  Kernels in Parzen 
  window density estimation (to be called density kernels) 
  are typically non-negative 
  symmetric functions integrating to 1 and 
  satisfying some conditions, while 
  kernels in RKHS are positive definite functions, 
  which are not necessarily positive.
  There are, of course, kernels that are both.  
  We will briefly review both, enough to review
  some modern results in two emerging fields, 
  density embedding and distance correlation.  
  Density  embedding begins with 
  an RKHS and a  sample from
  a density of interest and results
  in a class of density estimates which include
  Parzen window estimates. These  
  estimates are elements of the RKHS 
  {\it so one has 
  a metric for determining pairwise distances 
  between densities, namely the RKHS norm}. 
  This enlarges  the class of 
  familiar distance measures between densities 
  (e. g. Hellinger distance, Bhattacharyya 
  distance, Wasserstein distance, etc.) 
  Given  pairwise distances 
  between densities, we  then describe  how these 
  pairwise distances  are  used to include 
  sample densities 
  {\it as  attributes} in statistical learning 
  models such as Smoothing Spline ANOVA   
  (SS-ANOVA) models, which include  penalized likelihood
  methods and (nonparametric) 
  SVM's. Thus Manny's foundational
  work in two seemingly diverse areas come together to 
  add another feature to the statistician's tool kit.

\subsection{Two kinds of kernels}\marginpar{twokinds}
We now discuss the two kinds of kernels, those used in density 
estimation, and those that characterize an RKHS. 
Our goal is to show how sample densities possessed by 
subjects in RKHS-based prediction models  can be treated 
{\it as attributes} in these models.

\subsection{Parzen density kernels}
\marginpar{parden}
Let $X_1, X_2, \dots,  X_n$ be a random sample from some
(univariate) density $f(x), x \in (-\infty, \infty)$. The kernel
density estimates of Manny's seminal 1962 paper  
\cite{parzen:1962d} (paraphrasing slightly) 
are of the form 
\beq
f_n(x) = \frac{1}{nh}\sum_{j=1}^n K\left(\frac{x-X_j}{h}\right),
\eeq
where $K(y)$ is non-negative, 
\begin{equation}
\begin{split}
\sup_{-\infty <y< \infty}  K(y) &< \infty, \\
\int_{-\infty}^{\infty} K(y)&= 1,\\
\lim_{y \rightarrow \infty} |yK(y)| &= 0,
\end{split}
\end{equation}
and, letting $h = h(n),$ 
\beq
\lim_{n \rightarrow \infty} h(n) = 0.
\eeq

This landmark paper explores in detail the properties
of these density estimates, and gives a table of 
a number of $K$ that satisfy the requirements. 
Looking at the table of the $K$ 
and their Fourier transforms reveals that several but not
all are also positive definite. 

\subsection{RKHS kernels
}\marginpar{rkhs}
Manny was likely the first statistician to  
seriously introduce RKHSs to statisticians, certainly highly influential, 
see \cite{parzen:1962e,parzen:1963a,parzen:1970}.
As  a graduate student and postdoc at Stanford
from 1962-1967 I learned about RKHS directly
from Manny's lectures. 
Later the rich and beautiful results in \cite{parzen:1970} 
were highly influential in my own
life when work on splines at Madison rang a bell that
splines were a prototype of a vast class of 
nonparametric modeling 
problems  
that could be solved by RKHS methods, 
see \cite{kimeldorf:wahba:1971}.

Let $\cH_K$ be an RKHS 
of functions on a domain ${\cal{T}}$. 
Then
there exist a unique positive definite function
$K(s,t), s,t \in {\cal{T}}$ associated with $\cH_K$.
Conversely, let ${\cal{T}}$ be a domain on which 
a positive definite 
kernel function, $K(s,t), s,t \in {\cal{T}}$ 
 can be defined. Then there exists
 a unique RKHS $\cH_K$ 
 with $K$ as its reproducing kernel. This means 
 the following: Let $K_s(t) \equiv K(s,t)$ be considered
 as a function of $t$ for each fixed $s$. Then, 
 letting $<\cdot, \cdot>$ be the inner product in 
 $\cH_K$, for $f \in \cH_K$ we have $<f,K_s> = f(s)$, 
 and $<K_s, K_t> = K(s,t)$. 
 The square distance between $f$  and $g$ is 
 denoted  as $||f-g||^2_{\cH_K}$, where $||\cdot||^2_{\cH_K}$
 is the square norm in $\cH_K$.
 As a special case,  
 if $s,t \in {\cal{T}}$, then 
 the squared distance between $s$ and $t$ can be 
 taken as  
 $||K_s -K_t ||^2_{\cH_K} = K(s,s) -2K(s,t) +K(t,t)$.  
 We will be using the fact that  $K$ encodes 
 pairwise distances. We note that tensor sums and 
 products of positive definite functions are 
 positive definite functions and have associated
 RKHS as tensor sums and products of the corresponding 
 component RKHS, see
 \cite{aronszajn:1950} and the references cited
 below for examples.
\subsection{Smoothing Spline ANOVA models}\marginpar{rkhs2}
Basic references for SS-ANOVA models are
\cite{gu:2002} and \cite{wang:2011}, both describe 
software in the  R collection. Numerous applications include
\cite{gao:wahba:klein:klein:2001,lin:wahba:xiang:gao:2000,
wahba:wang:gu:klein:klein:1995}

Let $\cT^{(\al)}, \al = 1, \dots, d$ be $d$  domains 
with members $t_\al \in \cT^{(\al)}$. 
Let 
\[
t = (t_1,\dots, t_d) \in \cT^{(1)} \times \dots \times \cT^{(d)}  = \cT.
\] 
With each domain we have a positive definite function 
and associated RKHS. Now let $\cH_K$ be the 
tensor product of 
the $d$ RKHSs. Its RK is then the tensor product of the 
component RK's. With some conditions, including that 
the constant function is in each component space and there 
is an averaging operator in which the constant function
averages to 1, 
then for $f 
\in \cH_K$ an ANOVA decomposition %
of $f$ of the form  
\begin{equation}\label{form}
f(t_1, \cdots , t_d) = 
\mu + 
\sum_{\al} f_{\al}(t_\al) + \sum_{\al\be}f_{\al\be}(t_{\al}, t_{\be})
+ \cdots 
\end{equation}
can always be defined. 
Then a regularized kernel estimate is the 
solution to the problem

\begin{equation}
\min _{f \in \cH_K} \sum_{i=1}^n {\cal{C}}(y, f) + 
\lambda J(f),  
\end{equation}
where  ${\cal{C}}(y,f)$ relates to 
fit to predict $y$ from $f$, for example 
a Gaussian or Bernoulli log likelihood, or a hinge function (Support 
Vector Machine), and  
$J(f)$ is a square norm or seminorm 
in $\cH_K$.
Given this model for $f$ (generally truncated as warranted)
this provides a method for combining heterogenous 
domains (attributes) in a regularized prediction 
model. Note that nothing has been assumed about the 
domains, other than that a positive definite function
can be defined on them. 
We sketch an outline of facts relating to SS-ANOVA models, 
partly to set up notation to facilitate our goal
of demonstrating how sample densities may be treated 
as {\it attributes} in conjunction with SS-ANOVA models. 

The choice of kernel class for each variable may be an 
issue in practice and may be specific to the 
particular issue and data at hand. 
Once the kernel form has been chosen, 
the tuning parameter $\lambda$ in Equation (\ref{form})
along with other tuning parameters hidden in $J(f)$ 
must be chosen and can be important. 
We are omitting any discussion of 
these issues here, but  applications papers 
referenced below  discuss choice of tuning parameters.

Note 
that we use the same symbol $K$ for density kernels, positive 
definite functions and positive definite matrices.  


\marginpar{rkhs3}
Let $d\mu_{\al}$ be a probability measure on 
$\cT^{(\al)}$ and define the averaging operator 
$\cE_{\al}$ on $\cT$ by
\begin{equation}
({\cE}_{\al}f)(t)=
\int_{  {\cal{T}}^{(\al)}  }
f(t_1,\dots,t_d)
d\mu_{\al}(t_\al).
\end{equation}
Then the identity operator can be decomposed as
\[ 
I = \prod_{\al} (\cE_{\al} + (I-\cE_{\al})) =
\prod_{\al} \cE_{\al} + \sum_{\al}(I-\cE_{\al})\prod_{\be \neq \al} \cE_{\be}+ 
\] 
\[
 \sum_{\al < \be} (I-\cE_{\al})(I-\cE_{\be})
\prod_{\ga \neq \al,\be}\cE_{\ga}
+\cdots+ \prod_{\al}(I - \cE_{\al}),
\]
giving 
\[ 
\mu  = (\prod_{\al}\cE_{\al})f, ~~~ 
f_{\al} = ((I-\cE_{\al})\prod_{\be
\neq \al}\cE_{\be})f  ~~~~~~~
\] 
\[
f_{\al\be} = ((I -
\cE_{\al})(I-\cE_{\be})\prod_{\ga \neq \al,\be}\cE_{\ga})f \dots
\]

Further details
in the RKHS context may be found in 
\cite{gu:wahba:1993,wahba:1990,wahba:wang:gu:klein:klein:1995}.
\marginpar{rkhs4}
The idea behind SS-ANOVA  
is to construct an RKHS 
$\cH$  of functions on $\cT$ 
as the tensor product of RKHS on each $\cT^{(\al)}$ that
admit an ANOVA decomposition. 
Let $\cH^{(\al)}$ be an RKHS of functions on $\cT^{(\al)}$ with
$
\int _{\cT^{(\al)}} f_{\al} (t_{\al}) d\mu_{\al}(t_{\al}) = 0
$
and let  $[ 1^{(\al)}]$ be the one dimensional
space of constant functions on $\cT^{(\al)}$. 
Construct the RKHS  $\cH$ as 
\[ 
\cH  =  \prod _{\al=1}^d ([ 1^{(\al)}]  \oplus  \CH^{(\al)}  )
\]
\begin{equation}\label{halpha}
=  [1] \oplus \sum_{\alpha} \CH^{(\al)} \oplus 
      \sum_{\al < \be} [\CH^{(\al)} \otimes \CH^{(\be)} ] 
      \oplus \cdots,
\end{equation}
where $[1]$ denotes the constant functions on $\cT$. Then 
$f_{\alpha} \in \cH^{(\al)}, f_{\al\be} \in [\cH^{(\al)} \otimes \cH^{(\be)}]$ 
and so forth, where the series will usually be truncated at some point. 
Note that the usual ANOVA side conditions hold here.
\subsection{Pairwise distances in data analysis}
\marginpar{pairwise}

\subsubsection{Regularized Kernel Estimation}
Interesting examples of pairwise distances 
occur in, for example, blast scores
\cite{lu:keles:wright:wahba:2005} which give a 
pairwise dissimilarity
between pairs of protein sequences.
The blast score
pairwise dissimilarities are not a real distance, 
but they can be embedded (approximately) 
in a Euclidean space using Regularized Kernel Estimation 
(RKE)
\cite{lu:keles:wright:wahba:2005}. 

For a given $n \times n$ dimensional positive definite 
matrix $K$,  the pairwise distance that it induces is
$\hat{d}_{ij}=  K(i,i) +K(j,j)-2K(i,j)=B_{ij}\cdot K$, where
$K(i,j)$ is the $(i,j)$ entry of $K$ and
$B_{ij}$ is a symmetric $n \times n$ matrix 
with all elements
$0$ except $B_{ij}(i,i)=B_{ij}(j,j)=1$,
$B_{ij}(i,j)=B_{ij}(j,i)=-1$.  
The RKE 
problem is as follows: Given observed data $d_{ij}$ find
$K$  to 
\begin{equation} \label{rke:l1}
\min_{K\succeq0}\sum_{(i,j)\in\Omega}|d_{ij}-B_{ij}\cdot K|+
\lambda \, \mbox{trace}(K).
\end{equation}
$\Omega$ may be all pairs, or a connected subset. 

The data may be noisy/not Euclidean, but the RKE provides 
a (non-unique) embedding of the $n$  objects into an $r$-dimensional
Euclidean space as follows: Let the spectral decomposition 
of $K$ be $\Gamma \Lambda\Gamma^T$. The largest $r$  
eigenvalues and 
eigenvectors of  $K$ are retained to give the $n \times r$ matrix
$Z = \Gamma_r \Lambda_r^{1/2}$. We let the $i$th row of $Z$, 
an element of  $R^r$,
be the pseudo-attribute of the $i$th  subject. 

Thus  each subject  may be identified with an $r$-dimensional 
pseudo attribute, where the pairwise distances between the 
pseudo attributes respect (approximately, depending on $r$)
the original pairwise distances. 
Even if the original pairwise distances may be Euclidean, 
the RKE may be used as a dimension reduction procedure 
where the original pairwise distances have been obtained
in a much larger
space (e. g.  an infinite dimensional RKHS). 
The rank $r$ may be chosen to retain, say, $95\%$  
of the trace,  by examining an eigensequence plot for 
a sharp drop off, or maximizing the predictability in 
a supervised learning model. Note that if used in a predictive 
model it is necessary to know how a ``newbie" fits in; this is
discussed in \cite{lu:keles:wright:wahba:2005}.  

In the blast scores  example four well separated
clusters  
of known  proteins were readily evident in 
a three dimensional in-depth plot of the pseudo attributes,
and it could be seen that the multicategory 
support vector machine \cite{lee:lin:wahba:2004} would have 
classified the clusters nearly perfectly from  
these rank three pseudo attributes.  
Note that this embedding
is only unique up to a rotation, because rotating
the data set does not change the pairwise distance.
Therefore in fitting nonparametric models on the embedded
data only radial basis function (rbf) kernels may  be 
used, since they depend only on pairwise 
distances.   

\marginpar{pairwise1a}
Corrada Bravo {\it et al} \cite{corrada:wahba:lee:klein:2009} 
built a risk 
factor model consisting of an SS-ANOVA model 
with two genetic variables, life style attributes
and an additive term involving pairwise
distances of subjects
in pedigrees. The pedigree pairwise distances
were mapped into Euclidean space using RKE, and 
the Euclidean space of the  resulting pseudo
attributes used as the domain of an rbf 
based RKHS. 
The results were used to  examine the relative 
importance of genetic, lifestyle, and pedigree
information. It can be seen that this RKHS is not
treated as other terms in the SS-ANOVA model, as there are
no constant functions in the rbf based RKHS.  

Below we will see how sample densities can be embedded 
in an RKHS,  and pairwise distances and pseudo 
attributes obtained. Then the sample densities may be 
used in an SS-ANOVA model in the same way as in 
Corrada Bravo {\it et al}. 

\subsubsection{Pairwise distances reprised} \marginpar{pairwise2}
So, pairwise distances, either noisy or exact, 
may be included in information that can be built into
learning models. 
Applications of 
RK's in a variety of domains such as texts, 
images, strings and gene sequences, 
dynamical systems,  
graphs 
and structured
objects of various kinds 
have been defined. Recent examples include 
\cite{kondor:pan:2016} \cite{shen:wong:xiao:guo:2014}. 
We now proceed to examine pairwise distances
for sample densities. 

\subsection{Pairwise distances and kernel embedding for 
densities} \marginpar{embed}


Many  definitions of pairwise distance
between densities have appeared in the literature, 
in the context of testing for equality, including
Wasserstein distance, 
Bhattacharyya distance, 
Hellinger distance,
Mahalanobis distance,
among others.

Smola {\it et al} 
\cite{smola:gretton:song:scholkopf:2007} 
proposed to embed distributions into an RKHS,
and, once this is done, pairwise distances 
between a pair of distributions can be taken as 
the RKHS norm of the difference 
between the two embedded distributions.  

Let $\cH_K$ be an RKHS of functions on 
${\cal{T}}$ with RK $K(s,t), s,t \in {\cal{T}}$.
Let 
$X_1, X_2, \cdots, X_k$ be an iid sample from 
some density $p_X$. A map from this sample 
to $H_K$ is given by 

\begin{equation}
 f_X(\cdot) = \frac{1}{k} \sum_{j=1}^k K(X_j,\cdot). 
\end{equation}

Given a sample from a possibly different  distribution, 
we have 
\begin{equation}
g_Y(\cdot) = \frac{1}{\ell} \sum_{j=1}^{\ell} K(Y_j, \cdot).
\end{equation}

It is required that  $K$  be universal, among other things,  
\cite{s:g:s:k:2012,srip:fukumizu:lanckriet:2011},
which guarantees  that two 
different distributions
will be mapped into two different elements 
of $\cH_K$. See also
p. 727 of \cite{gretton:borgwardt:rasch:scholkopf:2012}.

\marginpar{embed2}
The pairwise distances between these two samples can be taken
as 
\newline
$\Vert f_X -g_Y\Vert_{\cal{H}_{K}}$
see \cite{srip:fukumizu:lanckriet:2011}, where
\begin{equation}
\Vert f_X -g_Y\Vert_{\cal{H}_K}= 
\frac{1}{k^2}\sum_{i, j = 1}^k K(X_i, X_j) 
+ \frac{1}{\ell^2} \sum_{i,j=1}^\ell K(Y_i, Y_j)
-\frac{2}{kl} \sum_{i=1,j=1}^{k,\ell}K(X_i,Y_j),
\end{equation}
thus providing a distance measure for each 
universal kernel to the other pairwise distances already
noted.  Note that if $K$ is a nonnegative, bounded 
radial basis function, then (up to scaling) we have mapped $f_X$ and $g_Y$ into Parzen type density estimates (!). The univariate version of a Gaussian rbf appears in Table 1 of 
Parzen \cite{parzen:1962d}.  

\marginpar{zhou}
Zhou {\it et al} 
\cite{zhou:ravi:ithapu:johnson:2016}
used pairwise embedding to consider
samples from two different data sources. They only
observed transformed versions $h(X_j), j = 1,2, \dots, k$
and $g(Y_j), j = 1,2 \dots, \ell$ 
for some known function class containing  
$h(\cdot)$ and $g(\cdot)$. The goal was to perform
a statistical test whether the two sources are
the same while removing the distortions induced
by the transformations.

\marginpar{embed3}
We already noted how Corrada Bravo {\it et al} 
\cite{corrada:wahba:lee:klein:2009}
used pairwise distances between pedigrees to include
pedigree information as an additive term in an 
SS-ANOVA model. Now, suppose we have a study where
subjects have various attributes, including a sample
density for each. One such example can be seen in 
\cite{nazapour:al-timemy:bugmann:jackson:2013}. 
Now  that we now have pairwise distances between pairs
of the sample densities, the densities can be included 
in an SS-ANOVA model as an additive term, using
the same approach as in \cite{corrada:wahba:lee:klein:2009}.

\subsection{Is density correlated with other variables?}
\marginpar{denscorr}

Distance Correlation (DCOR) \cite{szekely:rizzo:2009} is
key to an important area of recent research 
that uses pairwise distances only, to estimate a  
correlation-like quantity which behaves much 
like the Pearson 
correlation in the case of Gaussian variables, 
but provides a fully nonparametric test of independence
of two random variables. 
See \cite{szekely:rizzo:2009,szekely:rizzo:bakirov:2007}.
Recent contributions in the area include 
\cite{szekely:rizzo:2014}. 

For a random sample $(X,Y)=\{(X_k, Y_k): k=1,...,n\}$ of $n$ iid
random vectors $(X,Y)$ from the joint distribution of random vectors $X$ in $\mathrm{R}^p$ and $Y$ in $\mathrm{R}^q$, the Euclidean distance matrices $(a_{ij})=(|X_i-X_j|_p)$ and $(b_{ij})=(|Y_i-Y_j|_q)$ are computed. Define the 
{double centering distance matrices}
\begin{displaymath}
A_{ij}=a_{ij}-{\overline{a}_{i\cdot}}-{\overline{a}_{\cdot j}}+{\overline{a}_{\cdot\cdot}},\hspace{0.3cm}i,j=1,\dots,n,
\end{displaymath}
where
\begin{displaymath}
{\overline{a}_{i\cdot}}=\frac{1}{n}\sum_{j=1}^na_{ij},\hspace{0.3cm}{\overline{a}_{\cdot j}}=\frac{1}{n}\sum_{i=1}^na_{ij},\hspace{0.3cm}{\overline{a}_{\cdot\cdot}}=\frac{1}{n^2}\sum_{i,j=1}^na_{ij},
\end{displaymath}
similarly for $B_{ij}=b_{ij}-{\overline{b}_{i\cdot}}-{\overline{b}_{\cdot j}}+{\overline{b}_{\cdot\cdot}},\hspace{0.3cm}i,j=1,...,n$.\\
The sample distance covariance $\mathcal{V}_n(X,Y)$ is defined by
\begin{displaymath}
\mathcal{V}_n^2(X,Y)=\frac{1}{n^2}\sum_{i,j=1}^nA_{ij}B_{ij}.
\end{displaymath}
The sample {distance correlation} $\mathcal{R}_n(X,Y)$ (DCOR) is defined by

\begin{equation}
\mathcal{R}_n^2(X,Y)=\begin{cases}\displaystyle \frac{\mathcal{V}_n^2(X,Y)}{\sqrt{\mathcal{V}_n^2(X)\mathcal{V}_n^2(Y)}}, & \mathcal{V}_n^2(X)\mathcal{V}_n^2(Y)>0; \\
0, &\mathcal{V}_n^2(X)\mathcal{V}_n^2(Y)=0,\end{cases} \notag
\end{equation}
where the sample distance variance is defined by
\begin{displaymath}
\mathcal{V}_n^2(X)=\mathcal{V}_n^2(X,X)=\frac{1}{n^2}\sum_{i,j=1}^nA_{ij}^2.
\end{displaymath}
Distribution of the sample distance correlation under the 
null hypothesis of independence is easily found by scrambling
the data.

Kong {\it et al} \cite{kong:klein:klein:lee:2012} used 
DCOR and SS-ANOVA to assess
associations of familial relations and
lifestyle factors, diseases and mortality, 
by examining the strength of the rejection
of the null hypothesis of independence.  
Later, \cite{kong:wang:wahba:2015} used distance 
covariance 
as a greedy variable selector for learning a  
model with an extremely large number of 
candidate genetic variables.

\subsection{Including densities as attributes in an SS-ANOVA
model}

Suppose you have a population, \marginpar{denscorr2} 
each member having a (personal)  
sample density   and several other attributes, and you 
find using DCOR  that the individual sample densities 
are correlated with 
another variable in the model. The way to think about 
this is, when 
densities are close, so is the other variable, and 
vice versa. Interacting terms in the SS-ANOVA model
which include an rbf for the density RKHS can be included:
As in \cite{corrada:wahba:lee:klein:2009}, the densities 
are to
be embedded in some (generally infinite 
dimensional) rbf based  RKHS, and pairwise distances 
in this RKHS are determined.
RKE is then used to obtain pseudo attributes, which
are $r$ dimensional vectors,  and a second
rbf based
RKHS is chosen to model functions
of the pseudo attributes. 
The dimension $r$ of the pseudo attributes
can be controlled by the tuning parameter in 
the RKE.  
As noted earlier, the 
rbfs over $R^r$ do not in general contain a 
constant function, so they are treated a little
differently than the function spaces in the 
SS-ANOVA model that do. 
However,  tensor product  spaces consisting of the density 
RKHS $\cH^{(dens)}$ and other RKHS in the 
SS-ANOVA model after 
they have
been stripped of their constant functions may 
clearly be added to the model --- for example, 
suppose the density variable is correlated with 
the $\alpha$ variable, then 
$[\cH^{(\alpha)}\otimes \cH^{(dens)}]$ 
can be added to the model in 
Equation (\ref{halpha}), and  similarly for 
higher order interactions. 

\subsection{We have come full circle}\marginpar{fullcir}

So, we have come full circle. Manny proposed and 
investigated the properties of 
of Parzen kernel 
density estimates. Then Manny initiated 
an investigation into the various properties 
and importance of 
RKHS in new statistical methodology, 
and inspired me and many 
others to study these wonderful objects. 
So now we are 
able to include kernel density estimates
as attributes in SS-ANOVA models based on RKHS,
a modeling approach whose foundation lies in 
two of Manny's
major contributions to Statistical Science: 
density estimates, and Reproducing Kernel
Hilbert Spaces !

\section{Summary}\marginpar{summary}
In summary, I have been blessed to be one of 
Manny's students and  lifelong friends, 
and inspired by his path breaking work. He 
is terribly missed.

\bibliographystyle{plain} 
\bibliography{bib}

\begin{thebibliography}{10}

\bibitem{aronszajn:1950}
N.~Aronszajn.
\newblock Theory of reproducing kernels.
\newblock {\em Trans. Am. Math. Soc.}, 68:337--404, 1950.

\bibitem{corrada:wahba:lee:klein:2009}
H.~C. Bravo, K.~Lee, B.~E.~K. Klein, R.~Klein, S.~Iyengar, and G.~Wahba.
\newblock Examining the relative influence of familial, genetic, and
  environmental covariate information in flexible risk models.
\newblock {\em Proceedings of the National Academy of Sciences},
  106(20):8128--8133, 2009.

\bibitem{gao:wahba:klein:klein:2001}
F.~Gao, G.~Wahba, R.~Klein, and B.~Klein.
\newblock Smoothing spline {ANOVA} for multivariate {B}ernoulli observations,
  with applications to ophthalmology data, with discussion.
\newblock {\em J. Amer. Statist. Assoc.}, 96:127--160, 2001.

\bibitem{gretton:borgwardt:rasch:scholkopf:2012}
A.~Gretton, K.~Borgwardt, M.~Rasch, B.~Scholkopf, and A.~Smola.
\newblock A kernel two-sample test.
\newblock {\em J. Machine Learning Research}, 13:723--773, 2012.

\bibitem{gu:2002}
C.~Gu.
\newblock {\em Smoothing Spline ANOVA Models}.
\newblock Springer, 2002.

\bibitem{gu:wahba:1993}
C.~Gu and G.~Wahba.
\newblock Smoothing spline {ANOVA} with component-wise {B}ayesian ``confidence
  intervals''.
\newblock {\em J. Computational and Graphical Statistics}, 2:97--117, 1993.

\bibitem{hannan:1967}
E.~Hannan.
\newblock The estimation of a lagged regression relation.
\newblock {\em Biometrika}, 54:409--418, 1967.

\bibitem{kimeldorf:wahba:1970a}
G.~Kimeldorf and G.~Wahba.
\newblock A correspondence between {B}ayesian estimation of stochastic
  processes and smoothing by splines.
\newblock {\em Ann. Math. Statist.}, 41:495--502, 1970.

\bibitem{kimeldorf:wahba:1970b}
G.~Kimeldorf and G.~Wahba.
\newblock Spline functions and stochastic processes.
\newblock {\em Sankya Ser. A}, 32, Part 2:173--180, 1970b.

\bibitem{kimeldorf:wahba:1971}
G.~Kimeldorf and G.~Wahba.
\newblock Some results on {T}chebycheffian spline functions.
\newblock {\em J. Math. Anal. Applic.}, 33:82--95, 1971.

\bibitem{kondor:pan:2016}
R.~Kondor and H.~Pan.
\newblock The multiscale {L}aplacian graph kernel.
\newblock In {\em NIPS Proceedings 2016}. Neural Information Processing
  Society, 2016.

\bibitem{kong:klein:klein:lee:2012}
J.~Kong, B.~Klein, R.~Klein, K.~Lee, and G.~Wahba.
\newblock Using distance correlation and {S}moothing {S}pline {ANOVA} to assess
  associations of familial relationships, lifestyle factors, diseases and
  mortality.
\newblock {\em PNAS}, pages 20353--20357, 2012.
\newblock PMCID: 3528609.

\bibitem{kong:wang:wahba:2015}
J.~Kong, S.~Wang, and G.~Wahba.
\newblock Using distance covariance for improved variable selection with
  application to learning genetic risk models.
\newblock {\em Statistics in Medicine}, 34:1708--1720, 2015.
\newblock PMCID: PMC 4441212.

\bibitem{lee:lin:wahba:2004}
Y.~Lee, Y.~Lin, and G.~Wahba.
\newblock Multicategory support vector machines, theory, and application to the
  classification of microarray data and satellite radiance data.
\newblock {\em J. Amer. Statist. Assoc.}, 99:67--81, 2004.

\bibitem{lin:wahba:xiang:gao:2000}
X.~Lin, G.~Wahba, D.~Xiang, F.~Gao, R.~Klein, and B.~Klein.
\newblock Smoothing spline {ANOVA} models for large data sets with {B}ernoulli
  observations and the randomized {GACV}.
\newblock {\em Ann. Statist.}, 28:1570--1600, 2000.

\bibitem{lin:wahba:zhang:lee:2002}
Y.~Lin, G.~Wahba, H.~Zhang, and Y.~Lee.
\newblock Statistical properties and adaptive tuning of support vector
  machines.
\newblock {\em Machine Learning}, 48:115--136, 2002.

\bibitem{lu:keles:wright:wahba:2005}
F.~Lu, S.~Keles, S.~Wright, and G.~Wahba.
\newblock A framework for kernel regularization with application to protein
  clustering.
\newblock {\em Proceedings of the National Academy of Sciences},
  102:12332--12337, 2005.
\newblock Open Source at www.pnas.org/content/102/35/12332, PMCID: PMC118947.

\bibitem{nazapour:al-timemy:bugmann:jackson:2013}
K.~Nazapour, A.~Al-{T}imemy, G.~Bugmann, and A.~Jackson.
\newblock A note on the probability distribution function of the surface
  electromyogram signal.
\newblock {\em Brain Res. Bull.}, 90:88--91, 2013.

\bibitem{parzen:1960}
E.~Parzen.
\newblock {\em Modern Probability Theory and its Applications}.
\newblock Wiley, 1960.

\bibitem{parzen:1961b}
E.~Parzen.
\newblock Regression analysis of continuous parameter time series.
\newblock In {\em Proceedings of the Fourth Berkeley Symposium on Mathematical
  Statistics and Probability}, pages 469--489, Berkeley, California, 1960.
  University of California Press.

\bibitem{parzen:1961}
E.~Parzen.
\newblock An approach to time series analysis.
\newblock {\em Ann. Math. Statist.}, 32:951--989, 1961.

\bibitem{parzen:1962e}
E.~Parzen.
\newblock Extraction and detection problems and {R}eproducing {K}ernel
  {H}ilbert {S}paces.
\newblock {\em J. SIAM Series A Control}, 1:35--62, 1962.

\bibitem{parzen:1962d}
E.~Parzen.
\newblock On estimation of a probability density function and mode.
\newblock {\em Ann. Math. Statist.}, 33:1065--1076, 1962.

\bibitem{parzen:1962b}
E.~Parzen.
\newblock {\em Stochastic Processes}.
\newblock Holden-Day, San Francisco, 1962.

\bibitem{parzen:1963a}
E.~Parzen.
\newblock Probability density functionals and reproducing kernel {H}ilbert
  spaces.
\newblock In M.~Rosenblatt, editor, {\em Proceedings of the Symposium on Time
  Series Analysis}, pages 155--169. Wiley, 1963.

\bibitem{parzen:1970}
E.~Parzen.
\newblock Statistical inference on time series by {RKHS} methods.
\newblock In R.~Pyke, editor, {\em Proceedings 12th Biennial Seminar}, pages
  1--37, Montreal, 1970. Canadian Mathematical Congress.

\bibitem{parzen:1974}
E.~Parzen.
\newblock Some recent advances in time series modeling.
\newblock {\em IEEE Trans. Automatic Control}, AC-19:723--730, 1974.

\bibitem{s:g:s:k:2012}
D.~Sejdinovic, A.~Gretton, B.~Sriperumbudur, and K.~Fukumizu.
\newblock Hypothesis testing using pairwise distances and associated kernels.
\newblock arXiv:1205.0411v2, 2012.

\bibitem{shen:wong:xiao:guo:2014}
H-J. Shen, H-S. Wong, Q-W. Xiao, X.~Guo, and S.~Smale.
\newblock Introduction to the peptide binding problem of computational
  immunology: New results.
\newblock {\em Foundations of Computational Mathematics}, pages 951--984, 2014.

\bibitem{smola:gretton:song:scholkopf:2007}
A.~Smola, A.~Gretton, L.~Song, and B.~Scholkopf.
\newblock A {H}ilbert space embedding for distributions.
\newblock In {\em Algorithmic Learning Theory, 18th International Conference},
  pages 13--31. Springer Lecture Notes in Artificial Intelligence, 2007.

\bibitem{srip:fukumizu:lanckriet:2011}
B.~Sriperumbudur, K.~Fukumizu, and G.~Lanckriet.
\newblock Universality, characteristic kernels and rkhs embedding of measures.
\newblock {\em J. Machine Learning Research}, 12:2389--2410, 2011.

\bibitem{szekely:rizzo:2009}
G.~Szekely and M.~Rizzo.
\newblock Brownian distance covariance.
\newblock {\em Ann. Appl. Statist.}, 3:1236--1265, 2009.

\bibitem{szekely:rizzo:2014}
G.~Szekely and M.~Rizzo.
\newblock Partial distance correlation with methods for dissimilarities.
\newblock {\em Ann. Statist.}, 42:2382--2412, 2014.

\bibitem{szekely:rizzo:bakirov:2007}
G.~Szekely, M.~Rizzo, and N~Bakirov.
\newblock Measuring and testing independence by correlation of distances.
\newblock {\em Ann. Statist.}, 35:2769--2794, 2007.

\bibitem{wahba:1968}
G.~Wahba.
\newblock On the distribution of some statistics useful in the analysis of
  jointly stationary time series.
\newblock {\em Ann. Math. Statist.}, 39:1849--1862, 1968.

\bibitem{wahba:1969a}
G.~Wahba.
\newblock Estimation of the coefficients in a multi-dimensional distributed lag
  model.
\newblock {\em Econometrica}, 37:398--407, 1969.

\bibitem{wahba:1975c}
G.~Wahba.
\newblock Optimal convergence properties of variable knot, kernel and
  orthogonal series methods for density estimation.
\newblock {\em Ann. Statist.}, 3:15--29, 1975.

\bibitem{wahba:1977b}
G.~Wahba.
\newblock Optimal smoothing of density estimates.
\newblock In J.~VanRyzin, editor, {\em Classification and Clustering}, pages
  423--458. Academic Press, 1977b.

\bibitem{wahba:1980c}
G.~Wahba.
\newblock Automatic smoothing of the log periodogram.
\newblock {\em J. Amer. Statist. Assoc.}, 75:122--132, 1980c.

\bibitem{wahba:1990}
G.~Wahba.
\newblock {\em Spline Models for Observational Data}.
\newblock SIAM, 1990.
\newblock CBMS-NSF Regional Conference Series in Applied Mathematics, v. 59.

\bibitem{wahba:2013}
G.~Wahba.
\newblock Statistical model building, machine learning and the ah-ha moment.
\newblock In X.~Lin {\em et al}, editor, {\em Past, Present and Future of
  Statistical Science}, pages 481--495. Committee of Presidents of Statistical
  Societies, 2013.
\newblock Available at http://www.stat.wisc.edu/\~{}wahba/ftp1/tr1173.pdf.

\bibitem{wahba:wang:gu:klein:klein:1995}
G.~Wahba, Y.~Wang, C.~Gu, R.~Klein, and B.~Klein.
\newblock Smoothing spline {ANOVA} for exponential families, with application
  to the {W}isconsin {E}pidemiological {S}tudy of {D}iabetic {R}etinopathy.
\newblock {\em Ann. Statist.}, 23:1865--1895, 1995.
\newblock {N}eyman Lecture.

\bibitem{wold:wahba:1975e}
G.~Wahba and S.~Wold.
\newblock Periodic splines for spectral density estimation: {T}he use of
  cross-validation for determining the degree of smoothing.
\newblock {\em Commun. Statist.}, 2:125--141, 1975.

\bibitem{wang:2011}
Y.~Wang.
\newblock {\em Smoothing Splines: Methods and Applications}.
\newblock Chapman \& Hall/CRC Monographs on Statistics \& Applied Probability,
  2011.

\bibitem{zhou:ravi:ithapu:johnson:2016}
H.~H. Zhou, S.~Ravi, V.~Ithapu, S.~Johnson, G.~Wahba, and V.~Singh.
\newblock Hypothesis testing in unsupervised domain adaptation with
  applications in {A}lzheimer's disease.
\newblock In D.~D. Lee, M.~Sugiyama, U.~V. Luxburg, I.~Guyon, and R.~Garnett,
  editors, {\em Advances in Neural Information Processing Systems 29}, pages
  2496--2504. Neural Information Processing Society, 2016.
\newblock PMCID: PMC 5754041.

\end{thebibliography}
\end{document}